\documentclass[amsmath,11pt,amssymb]{revtex4}

\usepackage{color}
\usepackage[]{graphicx}
\usepackage{latexsym}
\usepackage{natbib}
\usepackage{amsmath}
\usepackage[caption=false]{subfig}
\usepackage{multirow}
\usepackage{mathtools}
\usepackage{textcomp}
\usepackage{xr}
\usepackage{slashed}
\usepackage{comment}

\usepackage[pdftex,hyperfigures,breaklinks,colorlinks,citecolor=black,linkcolor=black]{hyperref} 

\newcommand{\be}{\begin{equation}}
\newcommand{\ee}{\end{equation}}
\newcommand{\bc}{\begin{cases}}
\newcommand{\ec}{\end{cases}}

\renewcommand{\figurename}{Fig.}

\DeclareMathOperator*{\argmin}{arg\,min}

\begin{document}

\title{Ranking species in complex ecosystems through nestedness maximization}

\author{Manuel Sebastian Mariani}
\affiliation{Institute of Fundamental and Frontier Sciences,  University of Electronic Science and Technology of China, Chengdu 610054, P.R. China}
\affiliation{ URPP Social Networks, University of Zurich, CH-8050 Zurich, Switzerland}
\author{Dario Mazzilli, Aurelio Patelli}
\affiliation{ Enrico Fermi Research Center, via Panisperna 89a, 00184, Rome, Italy}
\author{Flaviano Morone}
\affiliation{Department of Physics, New York University, New York, New York 10003, USA}

\begin{abstract}


Identifying the rank of species in a social or ecological 
network is a difficult task, since the rank of each species 
is invariably determined by complex interactions stipulated 
with other species. 
Simply put, the rank of a species is a function of the 
ranks of all other species through the adjacency matrix 
of the network. 
A common system of ranking is to order species in such 
a way that their neighbours form maximally nested sets, 
a problem called nested maximization problem (NMP). 
Here we show that the NMP can be formulated as an instance of 
the Quadratic Assignment Problem, one of the most important 
combinatorial optimization problem widely studied in computer 
science, economics, and operations research. We tackle the problem
by Statistical Physics techniques: we derive a set of self-consistent 
nonlinear equations whose fixed point represents the optimal rankings 
of species in an arbitrary bipartite mutualistic network, 
which generalize the Fitness-Complexity equations widely used in 
the field of economic complexity. 
Furthermore, we present an efficient algorithm to solve the 
NMP that outperforms state-of-the-art network-based metrics 
and genetic algorithms. 
Eventually, our theoretical framework may be easily generalized 
to study the relationship between ranking and network structure 
beyond pairwise interactions, e.g. in higher-order networks.
\end{abstract}

\maketitle

\section{Introduction}
Experience reveals that species forming complex ecosystems
are organized in hierarchies. The ranks of such species, 
namely their position in the hierarchy, are functions of 
the interactions encoded in the adjacency matrix of the 
ecological network. Under this assumption, 
the task of ranking species can be cast in the problem of 
finding a suitable permutation of the rows and columns 
of the adjacency matrix, and this problem is, fundamentally, 
a combinatorial one. Ranking rows and columns of the adjacency matrix 
has revealed the existence of nested structures: 
neighbors of low rank nodes are subsets of the 
neighbors of high rank nodes~\cite{atmar1993measure,bascompte2003nested,mariani2019nestedness}.
For example, nested patterns are found in the world 
trade, in which products exported by low-fitness 
countries constitute subset of those exported 
by high-fitness countries~\cite{tacchella2012new}.
%
In fragmented habitats, species found in the least 
hospitable islands are a subset of species in the 
most hospitable islands~\cite{atmar1993measure}.
%
Nestedness in real world interaction networks 
has captured cross-disciplinary interest for three 
main reasons. First, nested patterns are ubiquitous 
among complex systems, ranging from ecological 
networks~\cite{atmar1993measure, bascompte2003nested}  
and the human gut microbiome~\cite{cobo2022stochastic} 
to socioeconomic systems~\cite{tacchella2012new,konig2014nestedness} 
and online social media and collaboration networks~\cite{palazzi2019online,palazzi2021ecological}.
Second, the ubiquity of nested patterns have 
triggered intensive debates about the reasons 
behind the emergence of nestedness in mutualistic systems~\cite{suweis2013emergence,valverde2018architecture,maynard2018network,cai2020mutualistic} and socioeconomic networks~\cite{konig2014nestedness,palazzi2021ecological}.
Third, nestedness may have profound implications 
for the stability and dynamics of ecological and economic communities:
highly-nested rankings of the nodes have revealed vulnerable species 
in mutualistic networks~\cite{dominguez2015ranking} and competitive 
actors in the world trade~\cite{tacchella2018dynamical, sciarra2020reconciling}.

The ubiquity of nestedness and its implications in 
shaping the structure of biotas
have motivated the formulation of the nestedness 
maximization problem. 
This problem can be stated in the following way: 
find the permutation (i.e. ranking) of the rows and 
columns of the adjacency matrix of the network 
resulting in a maximally nested layout of the 
matrix elements.
Originally introduced by Atmar and 
Patterson~\cite{atmar1993measure}, 
the problem has been widely studied in ecology, 
leading to several algorithms for measuring  
the nestedness of a matrix, e.g. the popular 
nestedness temperature calculator and its variants \cite{atmar1993measure,rodriguez2006new,almeida2007nestedness,payrato2020measuring}. 
Yet many of these methods do not attempt to optimize the 
actual cost of a nested solution, but exploit some simple 
heuristic that is deemed to be correlated with nestedness. 
Other methods, e.g. BINMATNEST~\cite{rodriguez2006new}, 
do optimize a nestedness cost following a genetic algorithm, 
but lack the theoretical insight contained in an analytic 
solution to the problem.  
More generally, we lack a formal theory to derive the 
degree of nestedness of a network from the structure 
of the adjacency matrix and the ranking of the nodes. 

Here, we map the nestedness maximization problem 
onto the Quadratic Assignement Problem~\cite{koopmans1957assignment}, 
thereby tackling directly the problem of finding the optimal 
permutation of rows and columns that maximizes the 
nestedness of the adjacency matrix.  
%
In our formulation, the degree of nestedness is 
measured by a cost function over the space 
of all possible rows and columns permutations, 
whose global minimum corresponds to a matrix 
layout having maximum nestedness. 
Roughly speaking, the cost function is designed 
to reward permutations that move the maximum 
number of non-zero elements of the matrix in 
the upper left corner and to penalize those 
that move non-zero elements in the bottom right 
corner. 
Next, we set up a theoretical framework which allows 
us to obtain the mean field solution to the NMP as a 
leading order approximation and, in principle, calculate 
also next-to-leading order corrections. 

\section{Problem formulation}

\begin{figure}
\includegraphics[scale=0.6]{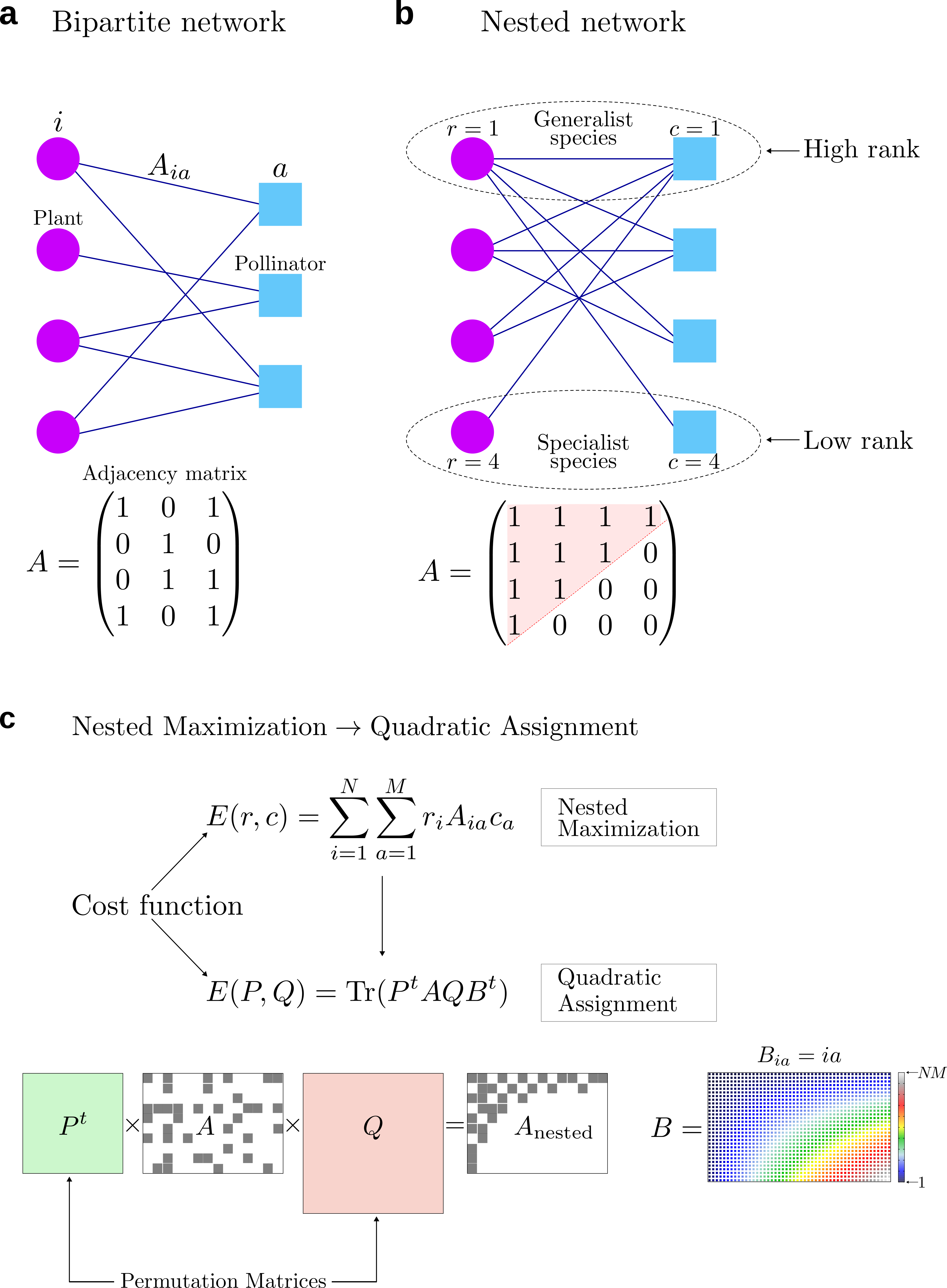}
\caption{{\bf Modeling of the Nested Maximization Problem}.
{\bf a}, A bipartite network models the interactions 
between, e.g., plants $i$, represented by purple circles, 
and pollinators $a$, represented by cyan squares, through 
the adjacency matrix $A$. The interaction is mutualistic, 
i.e. $A_{ia}=1>0$ if $i$ interacts with $a$ and $A_{ia}=0$ 
otherwise. 
{\bf b}, A nested network has a hierarchical 
structure wherein the neighbors of low rank nodes 
(the specialist species at the bottom) are a subset 
of the neighbors of high rank nodes (the generalists 
at the top). The rank of a node is encoded in 
the variables $r_i$ (for plants) and $c_a$ (for pollinators). 
Top rank nodes have $r=c-1$, while bottom ones have 
$r=c=4$. The adjacency matrix of a nested network shows 
a peculiar pattern with all non-zero entries clustered 
in the upper left corner. 
{\bf c}, Maximizing network nestedness amounts to minimize 
the cost function $E(r,c)$ over the ranking vectros $r$ 
and $c$, which, in turn, is equivalent to optimizing the 
cost $E(P,Q)$ with respect to the permutations matrices 
$P$ and $Q$. The optimal permutation matrices bring 
the adjacency matrix to its maximally nested form 
$P^tAQ=A_{\rm nested}$, which is complementary to the 
layout of matrix $B$. 
}
\label{fig:fig1}
\end{figure}

We consider bipartite networks where 
nodes of one kind, representing for example plants 
indexed by a variable $i=1,...,N$, can only be connected 
with nodes of another kind, e.g. pollinators indexed 
by another variable $a=1,...,M$, as seen in Fig.~\ref{fig:fig1}a.
We denote by $A_{ia}$ the element of the network's 
$N\times M$ adjacency matrix: 
$A_{ia}\neq 0$ if $i$ and $a$ are connected, and 
$A_{ia}=0$ otherwise. Besides connectivity, the 
adjacency matrix encodes the interaction strength 
between nodes such that whenever $i$ and $a$ are 
connected, the strength of their interaction is 
$A_{ia}=w_{ia}>0$.
A ranking of the rows is represented by a permutation 
of the integers $\{1,2,...,N\}$, denoted 
$r\equiv\{r_1,r_2,...,r_N\}$; a ranking of the columns 
is represented by a (different) permutation of the 
integers $\{1,2,...,M\}$, denoted $c\equiv\{c_1,c_2,...,c_M\}$. 
More precisely, the $r$ sequence arranges rows in 
ascending order of their ordinal rankings $r_i$ 
such that row $i$ is ranked higher than row $j$ if 
$r_i<r_j$.
Similarly, the $c$ sequence arranges columns 
such that column $a$ ranks higher than column 
$b$ if $c_a<c_b$.

To model the problem, one more concept is needed: network nestedness. 
Nestedness is the property whereby if $j$ 
ranks lower than $i$, than the neighbors 
of $j$ form a subset of the neighbors of $i$, 
as illustrated in Fig.~\ref{fig:fig1}b. 
Different rankings, i.e. different sequences $r$ 
and $c$, produce different nested patterns, that is, 
nestedness is a function of the rankings. 
Therefore, any cost (energy) function that seeks 
to quantify matrix nestedness must be a function 
of the rankings $r$ and $c$. 
The simplest energy function that does the job, 
aside from trivial cases (see Supplementary 
Information Sec.~\ref{sec:related_works}), is 
\begin{equation}
E(r,c)=\sum_{i=1}^N\sum_{a=1}^MA_{ia}r_ic_a\ .
\label{eq:energy1}
\end{equation}
The product $A_{ia}r_ic_a$ penalizes strong 
interactions between low-rank nodes, since 
they contribute a large amount to the cost function; 
thus, low rank nodes typically interact weakly. 
Strong interactions are only allowed between 
high rank nodes, because when $A_{ia}$ is large 
the product $A_{ia}r_ic_a$ can be made small by 
choosing $r_i$ and $c_a$ to be small. 
Furthermore, high rank nodes can have moderate 
interactions with low rank nodes, because the 
product $r_iA_{ia}c_a$ can be still relatively 
small when $r_i$ is large and $c_a$ is small (or 
viceversa) provided $A_{ia}$ is not 
too large (hence the name `moderate' interaction). 

The assumptions of our model are relevant to diverse 
scenarios where nestedness has been observed. 
In bipartite networks of countries connected to 
their exported products, we could interpret 
$r_i$ as the fitness of country $i$ and $c_a$ 
as the inverse of the complexity of product $a$.
In this scenario, high-energy links $r_iA_{ia}c_a$ 
represent the higher barriers faced by underdeveloped 
countries to produce and export sophisticated 
products~\cite{tacchella2012new}, whereas low-energy 
links represent competitive countries exporting 
ubiquitous products.
In mutualistic ecological networks, high-energy links 
represent the higher extinction risk for specialist 
pollinators to be connected with specialist plants, 
whereas low-energy links represent connections 
within the core of generalist nodes~\cite{bascompte2003nested} 
as depicted in Fig.~\ref{fig:fig1}b.

With this equipment, it should be clear that to maximize nestedness, we have to 
minimize the energy function in Eq.~\eqref{eq:energy1}. 
More precisely, nestedness maximization is the 
mathematical optimization problem in which we seek 
to find the optimal sequences $r^*$ and $c^*$ 
that minimize the energy function, i.e. 
$\min_{r,c}E(r,c) = E(r^*,c^*)$. 
Since the sequence $r$ is a permutation of the ordered 
sequence $\{1,2,...,N\}$, we can always write 
$r_i = \sum_{n=1}^NP_{in} n$, where $P$ is a $N\times N$ 
permutation matrix. Similarly, we can write 
$c_a = \sum_{m=1}^MQ_{am} m$ where $Q$ is a $M\times M$ 
permutation matrix. Therefore, the energy function, 
considered as a function of the permutation matrices 
$P$ and $Q$, can be rewritten in the form 
\begin{equation}
E(r,c) = E(P,Q) = {\rm Tr}\big(P^tAQB^t\big)\ , 
\label{eq:energy2}
\end{equation} 
where $B$ is a $N\times M$ matrix with entries 
$B_{ia}= ia$, as shown in Fig.~\ref{fig:fig1}c. 
In this language, the NMP is simply the problem of 
finding the permutations $P^*$ and $Q^*$ that minimize 
the energy function given by Eq.~\eqref{eq:energy2}, which mathematically reads
\begin{equation}
(P^*,Q^*)=\argmin_{P,\ Q}{E(P,Q)}\ .
\label{eq:problem}
\end{equation}
The geometric meaning of the optimal permutations 
$P^*$ and $Q^*$ is clear if we apply them to the 
adjacency matrix as $P^tAQ = A_{\rm nest}$ in that 
the nested structure in $A_{\rm nest}$ is visually 
manifest, as schematized in Fig.~\ref{fig:fig1}c. 
The optimization problem defined by Eqs.~\eqref{eq:energy2} 
and~\eqref{eq:problem} can be recognized as an instance of 
the Quadratic Assignment Problem (QAP) in the Koopmans-Beckmann 
form~\cite{koopmans1957assignment}, one of the most important 
problem in combinatorial optimization, that is known to be NP-hard. 
The formal mathematical mapping of the NMP onto an instance 
of the QAP represents our first most important result. 
Having formulated the NMP in the language of permutation 
matrices, we move next to solve it using a Statistical Physics 
approach.

\section{Solving the NMP with Statistical Physics}
Our basic tool to study the NMP is the partition 
function $Z(\beta)$ defined by
\begin{equation}
Z(\beta) = \sum_{P,\ Q} e^{-\beta E(P,Q)}\ , 
\end{equation}
where $\beta$ is an external control parameter, 
akin to the `inverse temperature' in the statistical 
physics language. 
The partition function $Z(\beta)$ provides a 
tool to determine the global minimum of the 
energy function via the limit 
\begin{equation}
E(P^*,Q^*)=-\lim_{\beta\to\infty}\frac{1}{\beta}\ln{Z(\beta)} 
\label{eq:groundstate}
\end{equation} 
Calculating the partition function may seem hopeless, 
since it requires to evaluate and sum up $N!M!$ terms. 
Nonetheless, the calculation is greatly simplified in the 
limit of large $\beta$, since we can evaluate $Z(\beta)$ 
via the steepest descent method. The strategy consists of 
two main steps. The first step is to work out an integral 
representation of $Z(\beta)$ of the form 
\begin{equation}
Z(\beta) = \int DX DY\ e^{-\beta F(X,Y)}\ ,
\label{eq:integralZ}
\end{equation}
where the integral is over the space of $N\times N$ 
doubly-stochastic (DS) matrices $X$ and $M\times M$ 
DS matrices $Y$, that converge onto permutation matrices 
$P$ and $Q$ when $\beta\to\infty$; and $F(X,Y)$ is an 
``effective cost function'' that coincides with 
$E(P,Q)$ for $\beta\to\infty$. 
The second step is to find the stationary points 
of $F(X,Y)$ by zeroing the derivatives  
$\partial F/\partial X = \partial F/\partial Y = 0$, 
resulting in a set of self-consistent equations for 
$X$ and $Y$, called saddle point equations. 
All steps of the calculation are explained in great 
detail in Supplementary Information~\ref{sec:SI_theory}. 
The resulting saddle point equations are given by 
\begin{equation}
\begin{aligned}
X_{ij} &= u_i \exp\Big[-\beta\big(AYB^t)_{ij}\Big] v_j ,\\
Y_{ab} &= \mu_a \exp\Big[-\beta\big(A^tXB)_{ab}\Big]\nu_b\ ,
\end{aligned}
\label{eq:sp1}
\end{equation}
where $u, v$ are $N$-dimensional vectors and 
$\mu,\nu$ are $M$-dimensional vectors determined 
by imposing that all row and column sums of $X$ and 
$Y$ are equal to 1. 
At this point we can exploit the specific form of matrix 
$B$, i.e. $B_{ia}=ia$, to further simplify Eqs.~\eqref{eq:sp1}. 
Specifically, we define the ``stochastic'' rankings $\rho_i$ and 
$\sigma_a$ as 
\begin{equation}
\rho_i = \sum_{k=1}^N X_{ik}\ k\ ,\ \ \ 
\sigma_a = \sum_{b=1}^M Y_{ab}\ b\ , 
\end{equation}
whereby we can cast Eqs.~\eqref{eq:sp1} in the following 
vectorial form (details in Supplementary Information~\ref{sec:SI_theory})
%
\begin{equation}
\begin{aligned}
\rho_i &= \frac{\sum_k k\ v_k\ e^{-\beta k\sum_a A_{ia}\sigma_a}}{\sum_k v_k\ e^{-\beta k\sum_a A_{ia}\sigma_a}}\ ,\\
\sigma_a &= \frac{\sum_c c\ \nu_c\ e^{-\beta c\sum_i A_{ia}\rho_i}}{\sum_c \nu_c\ e^{-\beta c\sum_i A_{ia}\rho_i}}\ ,
\end{aligned}
\label{eq:rankPropagation1}
\end{equation}
where the normalizing vectors $v$ and $\nu$ satisfy 
\begin{equation}
\begin{aligned}
\frac{1}{v_j} = 
\sum_i\Big[\sum_k\ v_k\   e^{-\beta (k-j)\sum_a A_{ia}\sigma_a}\Big]^{-1}\ ,\\ 
\frac{1}{\nu_b} = 
\sum_a\Big[\sum_c\ \nu_c\ e^{-\beta (c-b)\sum_i A_{ia}\rho_i}\Big]^{-1}\ .
\end{aligned}
\label{eq:rankPropagation2}
\end{equation}
Equations~\eqref{eq:rankPropagation1} and~\eqref{eq:rankPropagation2}  
represent our second most important result and, when interpreted 
as iterative equations, provide a simple algorithm to solve the NMP, 
whose implementation is discussed in detail in Supplementary 
Information~\ref{sec:algorithm}. 
Note that $\rho$ and $\sigma$ converge to the the 
actual ranking $r$ and $s$ for $\beta\to\infty$. 
However, in practice, we solve Eqs.~\eqref{eq:rankPropagation1} 
and~\eqref{eq:rankPropagation2} iteratively at finite 
$\beta$. 
Once we reach convergence, we estimate $r$ and $s$ by 
simply sorting the entries of $\rho$ and $\sigma$. We observe that 
larger values of $\beta$ give better results, i.e., lower values 
of the cost $E(r,s)$, as seen in Fig.~\ref{fig:fig2}a. 
A full discussion of convergence and bounds of our algorithm will 
be published elsewhere. Here, we test its performance by 
applying it to many real mutualistic networks and show that 
we obtain better results than state-of-the-art network metrics 
and genetic algorithms, as discussed next.

\section{Numerical results}
\begin{figure}
\includegraphics[scale=0.6]{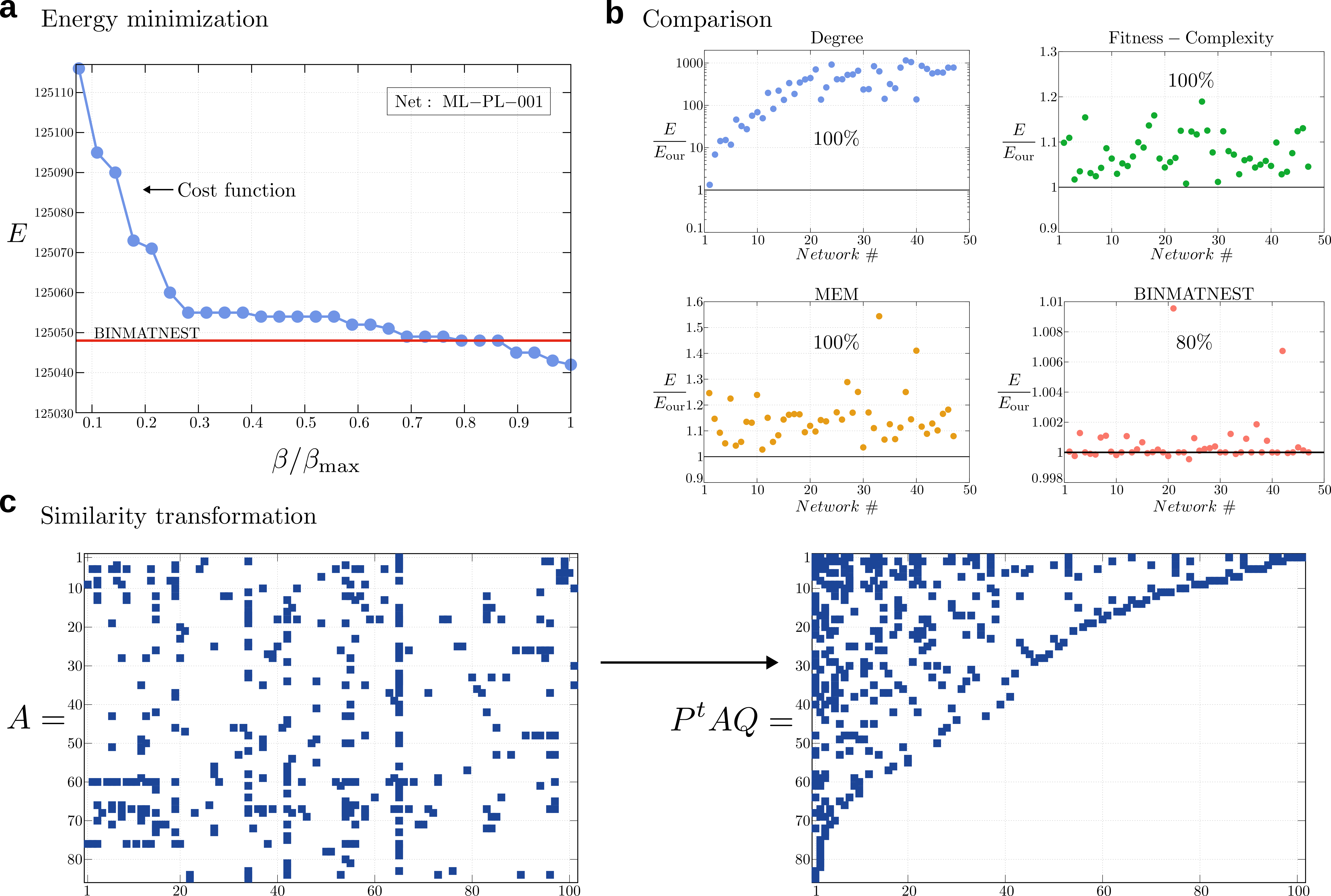}
\caption{{\bf Numerical solution and comparison with other methods}.
{\bf a}, Optimal cost $E(r,c)$ returned by our algorithm on the 
mutualistic network named {\it ML-PL-OO1} in the Web-of-Life database, 
for several choices of the parameter $\beta$. Larger values of $\beta$ 
give lower costs. In particular, for sufficiently large $\beta$ our 
algorithm returns a lower cost than the best off-the-shelf algorithm 
for nestedness maximization (BINMATNEST, red line). 
{\bf b}, Comparison of our algorithm with state-of-the-art methods 
in the literature: Degree (upper-left), Fitness-Complexity (upper-right), 
Minimal-Extremal-Metric (bottom-left) , and BINMATNEST (bottom-right). 
In each panel we plot the cost returned by each algorithm divided 
by the cost returned by our algorithm, denoted $E/E_{\rm our}$, for 
each network considered in this work. A value $E/E_{\rm our}>1$ means 
that our algorithm returns a better, i.e. lower, cost. We find that 
our algorithm returns a better cost in $100\%$ of the networks when 
compared to degree, FC, and MEM, and in $80\%$ of the networks when 
compared to BINMATNEST (see also Table~\ref{tab:table1}). 
{\bf c}, Similarity transformation applied to the adjacency matrix 
$A$ of network {\it ML-PL-OO1} that brings $A$ into its maximally 
nested form $P^tAQ$, where $P$ and $Q$ are the optimal permutation 
matrices constructed from the optimal ranking vectors $r^*$ and $s^*$.  
}
\label{fig:fig2}
\end{figure}

We apply our algorithm on 47 real mutualistic networks freely 
downloadable at~\url{https://www.web-of-life.es/}, whose filenames 
can be found in the first column of Table~\ref{tab:table1}. 
To standardize the comparison with existing methods, we binarize 
the adjacency matrices of the networks setting $A_{ij}=1$ if nodes 
$i$ and $j$ are connected and zero otherwise, thus ignoring the 
weights.
Despite this simplification, we like to emphasize that our 
algorithm can be applied, as is, to any mutualistic weighted 
network of the most general form. 
Then we run four different algorithms comprising: naive degree~\cite{araujo2010analytic}, 
fitness-complexity (FC)~\cite{tacchella2012new}, minimal extremal metric (MEM)~\cite{wu2016mathematics}, 
and BINMATNEST~\cite{rodriguez2006new}.
While BINMATNEST is the state-of-the-art algorithm in ecology for nestedness maximization~\cite{dormann2020using}, the effectiveness of the FC~\cite{lin2018nestedness,mazzilli2022fitness} and MEM~\cite{wu2016mathematics} has been proved in recent works in economic complexity, which also connected the FC to the Sinkhorn algorithm from optimal transport~\cite{sinkhorn1967concerning,marshall1968scaling,mazzilli2022fitness}.
We compare the value of the 
cost function $E(r,c)$ returned by each of the analyzed algorithms to the value returned by 
our algorithm (see Supplementary Information Sec.~\ref{sec:related_works} 
for implementation details). As shown in Fig.~\ref{fig:fig2}b, our algorithm finds a 
better (i.e. lower) cost than degree, FC, and MEM on $100\%$ of the networks. 
When compared to BINMATNEST, we find a better (or equal) 
minimum cost in $80\%$ of the instances, as seen in Fig.~\ref{fig:fig2}b and 
Table~\ref{tab:table1}. 

We conclude this section by showing an application of the 
similarity transformation that brings the adjacency matrix 
to its maximally nested form. 
We call $P$ and $Q$ the optimal permutations that solve the 
QAP in Eq.~\eqref{eq:problem} (details in Supplementary 
Information Sec.~\ref{sec:algorithm}) and we perform the 
similarity trasformation 
\begin{equation}
A\to P^tA Q\ ,
\end{equation}
which reveals the nested structure of the adjacency matrix 
shown in Fig.~\ref{fig:fig2}c.

\section{Conclusions}

In this work we introduced a cost function for the NMP 
in bipartite mutualistic networks. This formulation allowed 
us to recast the problem as an instance of the QAP, that we 
tackled by Statistical Physics techniques. 
In particular, we obtained a mean field solution by using 
the steepest-descent approximation of the partition 
function. The corresponding saddle-point equations depend on 
a single hyper-parameter (the inverse temperature $\beta$) and 
can be solved by iteration to find the optimal rankings of the 
rows and columns of the adjacency matrix that result in a maximally 
nested layout. 
We benchmarked our algorithm against other methods on several 
real ecological networks and showed that our algorithm outperforms 
the best existing algorithm in $80\%$ of the instances. 

We note that by changing the definition of the 
matrix $B$, i.e. using measures other than a sequence of ordinal 
numbers, one can repurpose our algorithm to rank rows and columns 
of a matrix according to other geometric patterns~\cite{morone2022clustering, de2018physical}. 
Therefore, the proposed framework holds promise for the effective 
detection of a wide range of network structural patterns beyond the 
nestedness considered here.
Finally, the present framework can be easily extended 
and applied to solve the ranking problem in networks 
with higher order interactions. For example, given the 
adjacency tensor $A_{ia\gamma}$ for a system with $3$-body 
interactions, we can define the energy function 
$E(P,Q,R)$ to be optimized over $3$ permutation 
matrices $P$, $Q$, and $R$ following exactly the same 
steps outlined in this paper for the case of pairwise 
interactions. This may be especially relevant in the world 
trade for ranking countries according to both exported and 
imported goods.

\begin{center}
\begin{tabular}{|c|c|c|c|c|c|c|c|c|c|}
\hline
Net & N & M & $||A||/NM$ & FC & DEG & MEM & BIT & OUR  \\ [0.5ex]
\hline\hline
M-PL-001 & 84 & 101 & 0.042551 & 137348 & 165930 & 155841 & 125048 & \textcolor{blue}{125042}  \\ 
\hline
M-PL-002  & 43 & 64 & 0.071221 & 37556 & 232827 & 38823 & \textcolor{blue}{33850} & 33858  \\ 
\hline
M-PL-003 & 36 & 25 & 0.090000 & 3927 & 55335 & 4220 & 3866 & \textcolor{blue}{3862}  \\ 
\hline
M-PL-004  & 12 & 102 & 0.136438 & 12082 & 176999 & 12274 & \textcolor{blue}{11672} & \textcolor{blue}{11672}  \\ 
\hline
M-PL-005  & 96 & 275 & 0.034962 & 885890 & 9040760 & 939937 & \textcolor{blue}{767320} & 767393  \\ 
\hline
M-PL-006  & 17 & 61 & 0.140791 & 6579 & 293503 & 6653 & \textcolor{blue}{6379} & \textcolor{blue}{6379}  \\ 
\hline
M-PL-007 & 16 & 36 & 0.147569 & 3109 & 98372 & 3210 & 3038 & \textcolor{blue}{3036}  \\ 
\hline
M-PL-008 & 11 & 38 & 0.253589 & 5654 & 148325 & 6153 & 5428 & \textcolor{blue}{5422}  \\ 
\hline
M-PL-009 & 24 & 118 & 0.085452 & 48398 & 2535535 & 50418 & 44559 & \textcolor{blue}{44556}  \\ 
\hline
M-PL-010  & 31 & 76 & 0.193548 & 103649 & 6714987 & 120773 & \textcolor{blue}{97454} & 97472  \\ 
\hline
M-PL-011  & 14 & 13 & 0.285714 & 970 & 46815 & 968 & \textcolor{blue}{943} & \textcolor{blue}{943}  \\ 
\hline
M-PL-012 & 29 & 55 & 0.090909 & 9948 & 1861449 & 10871 & 9460 & \textcolor{blue}{9449}  \\ 
\hline
M-PL-013  & 9 & 56 & 0.204365 & 4863 & 383760 & 4910 & \textcolor{blue}{4644} & \textcolor{blue}{4644}  \\ 
\hline
M-PL-014 & 29 & 81 & 0.076203 & 20106 & 4179783 & 20387 & 18830 & \textcolor{blue}{18827}  \\ 
\hline
M-PL-016 & 26 & 179 & 0.088526 & 122835 & 15019420 & 127784 & 111800 & \textcolor{blue}{111725}  \\ 
\hline
M-PL-017  & 25 & 79 & 0.151392 & 35393 & 10925775 & 37814 & \textcolor{blue}{32533} & 32534  \\ 
\hline
M-PL-018 & 39 & 105 & 0.093529 & 121642 & 19872497 & 124677 & 107023 & \textcolor{blue}{107022}  \\ 
\hline
M-PL-019 & 40 & 85 & 0.077647 & 56643 & 16872116 & 56890 & 48888 & \textcolor{blue}{48879}  \\ 
\hline
M-PL-020  & 20 & 91 & 0.104396 & 17037 & 6545141 & 17540 & \textcolor{blue}{16022} & \textcolor{blue}{16022}  \\ 
\hline
M-PL-022  & 21 & 45 & 0.087831 & 4339 & 1833172 & 4655 & \textcolor{blue}{4156} & 4158  \\ 
\hline
M-PL-023 & 23 & 72 & 0.075483 & 9513 & 6341662 & 9890 & 9098 & \textcolor{blue}{9011}  \\ 
\hline
M-PL-024  & 11 & 18 & 0.191919 & 803 & 103022 & 862 & \textcolor{blue}{755} & \textcolor{blue}{755}  \\ 
\hline
M-PL-025  & 13 & 44 & 0.250000 & 8148 & 1921580 & 8233 & \textcolor{blue}{7243} & \textcolor{blue}{7243}  \\ 
\hline
M-PL-026  & 105 & 54 & 0.035979 & 17998 & 16395570 & 56197 & \textcolor{blue}{17847} & 17855  \\ 
\hline
M-PL-027 & 18 & 60 & 0.111111 & 14188 & 5208823 & 14803 & 12644 & \textcolor{blue}{12633}  \\ 
\hline
M-PL-028 & 41 & 139 & 0.065626 & 126748 & 46897882 & 129783 & 113503 & \textcolor{blue}{113490}  \\ 
\hline
M-PL-029 & 49 & 118 & 0.059841 & 105634 & 46529364 & 114448 & 88825 & \textcolor{blue}{88805}  \\ 
\hline
M-PL-030 & 28 & 53 & 0.073450 & 15658 & 7451270 & 16284 & 13918 & \textcolor{blue}{13915}  \\ 
\hline
\end{tabular}
\end{center}

\begin{table}[!h]
\begin{center}
\begin{tabular}{|c|c|c|c|c|c|c|c|c|c|}
\hline
Net & N & M & $||A||/NM$ & FC & DEG & MEM & BIT & OUR  \\ [0.5ex]
\hline\hline
M-PL-031 & 48 & 49 & 0.066327 & 24134 & 14712154 & 28025 & 22418 & \textcolor{blue}{22409}  \\ 
\hline
M-PL-032  & 7 & 33 & 0.281385 & 1379 & 322338 & 1413 & \textcolor{blue}{1363} & \textcolor{blue}{1363}  \\ 
\hline
M-PL-033  & 13 & 34 & 0.319005 & 9718 & 2086383 & 10128 & \textcolor{blue}{8648} & \textcolor{blue}{8648}  \\ 
\hline
M-PL-034 & 26 & 128 & 0.093750 & 48523 & 37671897 & 49907 & 44993 & \textcolor{blue}{44938}  \\ 
\hline
M-PL-035  & 61 & 36 & 0.081056 & 19907 & 11775325 & 28663 & \textcolor{blue}{18565} & 18567  \\ 
\hline
M-PL-036  & 10 & 12 & 0.250000 & 465 & 64621 & 483 & \textcolor{blue}{452} & \textcolor{blue}{452}  \\ 
\hline
M-PL-037 & 10 & 40 & 0.180000 & 3543 & 1061073 & 3763 & 3346 & \textcolor{blue}{3342}  \\ 
\hline
M-PL-038  & 8 & 42 & 0.235119 & 3616 & 860044 & 3631 & \textcolor{blue}{3399} & \textcolor{blue}{3399}  \\ 
\hline
M-PL-039 & 17 & 51 & 0.148789 & 8400 & 6259559 & 8956 & 8065 & \textcolor{blue}{8050}  \\ 
\hline
M-PL-040  & 29 & 43 & 0.091419 & 8126 & 8906049 & 9676 & \textcolor{blue}{7739} & \textcolor{blue}{7739}  \\ 
\hline
M-PL-041 & 31 & 43 & 0.108777 & 12445 & 12353208 & 13463 & 11771 & \textcolor{blue}{11761}  \\ 
\hline
M-PL-042  & 12 & 6 & 0.347222 & 221 & 29225 & 298 & \textcolor{blue}{212} & \textcolor{blue}{212}  \\ 
\hline
M-PL-043  & 28 & 82 & 0.108885 & 46324 & 36103187 & 47058 & \textcolor{blue}{42156} & \textcolor{blue}{42156}  \\ 
\hline
M-PL-045 & 17 & 26 & 0.142534 & 1833 & 1291777 & 1941 & 1795 & \textcolor{blue}{1783}  \\ 
\hline
M-PL-046  & 16 & 44 & 0.394886 & 23365 & 12810171 & 25494 & \textcolor{blue}{22591} & 22592  \\ 
\hline
M-PL-047  & 19 & 186 & 0.120260 & 82943 & 46841210 & 84968 & \textcolor{blue}{77126} & \textcolor{blue}{77126}  \\ 
\hline
M-PL-048 & 30 & 236 & 0.094774 & 273971 & 144577341 & 284223 & 243852 & \textcolor{blue}{243771}  \\ 
\hline
M-PL-049 & 37 & 225 & 0.070871 & 255534 & 175524328 & 267224 & 226068 & \textcolor{blue}{226039}  \\ 
\hline
M-PL-050  & 14 & 35 & 0.175510 & 3467 & 2586805 & 3581 & \textcolor{blue}{3317} & \textcolor{blue}{3317}  \\ 
\hline
\end{tabular}
\caption{Numerical results on real mutualistic networks from the Web of Life database. 
First tab is the filename of the network as it appears in the database. Second and 
third tabs are the number of rows and columns, respectively. Fourth tab is the norm 
of the (binarized) adjacency matrix (sum of non zero entries) divided by $NM$. 
Last five tabs represent the minimum cost returned by, 
in order, Fitness-Complexity, Degree, Minimal Extremal Metric, BINMATNEST and our method. We highlight in blue the best 
result among these five methods. 
}
\label{tab:table1}
\end{center}
\end{table}

\newpage

\noindent
{\bf \large Data availability}
Data that support the findings of this study are publicly 
available at the Web of Life database at~\url{https://www.web-of-life.es/} 

\medskip

\noindent
{\bf \large Acknowledgments} 
This work was partially supported by AFOSR: Grant FA9550-21-1-0236. MSM acknowledges financial support from the URPP Social Networks at the University of Zurich, and the Swiss National Science Foundation, Grant 100013-207888.

\medskip

\noindent
{\bf \large Author contributions}
All authors contributed equally to this work.

\medskip

\noindent
{\bf \large Additional information}
Supplementary Information accompanies this paper. 

\medskip

\noindent
{\bf \large Competing interests} 

The authors declare no competing interests. 

\medskip

\noindent
{\bf Correspondence} should be addressed to F. M. at: fm2452@nyu.edu


\clearpage


\clearpage


\clearpage

\renewcommand{\figurename}{Supplementary Figure}
\renewcommand{\tablename}{Supplementary Table}

\centerline{ \bf \large Supplementary Information for:}

\centerline{\bf Ranking species in complex ecosystems through nestedness maximization}

\medskip

\centerline{Manuel Sebastian Mariani, Dario Mazzilli, 
Aurelio Patelli \& Flaviano Morone}

\tableofcontents

\clearpage

\section{Related Works}
\label{sec:related_works}

In this section we briefly review existing methods, models, 
and algorithms tackling the ranking and nestedness maximization 
problems.

\subsection{Ranking by degree}

The degree of a node is simply defined as its number of connections. 
It can be connected to a nestedness maximization problem as follows.
In Ref.~\cite{araujo2010analytic} the authors 
consider the following energy function
\begin{equation}
E(\vec{r},\vec{s})= \sum_{i,a}A_{ia}(r_i+s_a)\ .
\label{eq:SI_energy_araujo}
\end{equation}
The meaning of this energy function can be 
easily understood when $A_{ia}\in\{0,1\}$. 
In this case the sum can be rewritten as: 
$\sum_{ia}A_{ia}(r_i+s_a) = \sum_i k_ir_i + \sum_a k_as_a$, 
where $k_i$ and $k_a$ are the degrees 
(number of connections) of nodes $i$ and $a$, respectively.  
In the language of statistical physics 
the term $k_ir_i$ represents an interaction 
between the degrees of freedom $r_i$ and a 
local magnetic field $k_i$, whose intensity equals 
the node's degree. 
The stronger the magnetic field $k_i$ is, the 
lower the value of $r_i$ ought to be in order 
to minimize the product $r_ik_i$. 
This reasoning can be generalized to the case 
$A_{ia}\in\{0,w_{ia}\}$ upon changing the 
definition of the magnetic field from the node 
degree to the weighted node degree, the weights 
being the interaction strengths  $w_{ia}$. 
In both cases, the effect of this term is to 
assign high rank to nodes with high values 
of $k_i$ (or $k_a$ of course).

The non-interacting energy function defined 
in~\eqref{eq:SI_energy_araujo} is minimized by 
ranking the nodes according to their degree, and can be 
seen as an instance of the Linear Assignment Problem, 
whose solution can be found in polynomial time (in this 
case by simply sorting the degrees, so in $O(N\log N)$ 
operations).
Authors of Ref.~\cite{araujo2010analytic} only 
considered the rankings of nodes by degree, and 
they were interested in comparing the energy 
observed in empirical networks against that of 
idealized nested structures.
In our framework, we model the nestedness maximization 
problem by an energy function that couples the rows and 
columns' ranking positions, which can be seen as an 
instance of the Quadratic Assignment Problem~\cite{koopmans1957assignment}, 
which is known to be NP-hard, and thus there is no known 
algorithm that can find the optimal solution in polynomial 
time. 

\subsection{SpringRank}

Reference~\cite{de2018physical} considered an energy-based approach to rank nodes in directed 
weighted unipartite networks. They defined $A_{ij}$ 
as the number of interactions suggesting that $i$ is 
ranked above $j$, and they defined the SpringRank 
centrality as the vector $\vec{\eta}^*$ of real-valued 
scores that minimize the energy function
\begin{equation}
E(\vec{\eta})= \sum_{i,j}A_{ij}(\eta_i-\eta_j+1)^2\ .
\label{eq:SI_energy_debacco}
\end{equation}
The model reflects the assumption that if many directed 
interactions suggesting that $i$ is ranked above $j$ are 
observed, then the centrality of $i$ should be much larger 
than that of $j$.
Subsequently, the authors develop statistical inference 
techniques to infer the node-level SpringRank scores in 
empirical networks. 
Broadly speaking, their approach is conceptually related to ours as it defines the rankings of the nodes in terms of the minimum of an energy function that depends on the nodes scores and the network's adjacency matrix.
However, their ranking method focuses on directed weighted unipartite networks and it does not aim 
at maximizing the network nestedness, and therefore it won't be compared to the method 
presented in this work.

\subsection{BINMATNEST}

BINMATNEST~\cite{rodriguez2006new} can be considered as the state-of-the-art algorithm to maximize nestedness in ecology~\cite{dormann2020using}.
In fact, the algorithm minimizes the nestedness temperature~\cite{atmar1993measure}, a variable that is conceptually related to the nestedness energy defined in the main text.
The nestedness temperature $T$ quantifies the average distance of the adjacency matrix’s elements from the so-called isocline of perfect nestedness, which represents the separatrix between the empty and filled regions of a perfectly-nested matrix with the same density as the original matrix. We refer to \cite{rodriguez2006new} for details of the isocline determination and temperature calculation. Of course $T$ depends on the adjacency matrix $A$ as well as the permutation of its rows and columns. The dependence of $T$ on the ranking vectors is more complex than the nestedness energy function introduced here, and therefore, its optimization less amenable to analytic treatment. The genetic algorithm BINMATNEST bypasses the problem by relying on an iterative algorithm.

In BINMATNEST~\cite{rodriguez2006new}, a candidate solution is represented by the rankings’ vectors $r=\{r_1,r_2,…,r_N\}$ and $c=\{c_1,c_2,…,c_M\}$. One starts from a population of initial solutions, composed of the original matrix, solutions found with a similar algorithm as the original one by Atmar and Patterson~\cite{atmar1993measure}, and their mutations. From a well-performing candidate solution, an offspring of solution is created by selecting a second “parent” from the remaining solutions in the population, suitably combining the information from the two solutions, and eventually performing random mutations in the resulting child solution. Specifically, denote as $w$ the row ranking vector of a well-performing solution and $p$ the row ranking vector of its selected partner (the procedure is analogous for the column ranking vectors). The row ranking vector of the offspring solution, $o$, is set to $w$ with probability $0.5$, otherwise it is determined by a combination of $w$ and $p$ determined by the following algorithm~\cite{rodriguez2006new}:
\begin{itemize}
\item An integer $k\in\{1,\dots,N\}$ is selected uniformly at random.
\item We set $o_i=w_i$ for all $i\in\{1,\dots,k\}$.
\item For $i\in\{k+1,\dots,N\}$, if $p_i\notin \{w_i,\dots,w_k\}$, then we set $o_i=p_i$.
\item For $i\in\{k+1,\dots,N\}$, if $p_i\in \{w_i,\dots,w_k\}$, then the value of $o_i$ is chosen at random from all the unused positions.
\end{itemize}
As final step, a random mutation of ranking vector $o$ is performed by selecting at random $k_1,k_2\in\{1,\dots,N\}$ and performing a cyclical permutation of the elements $r_{k_1},\dots,r_{k_2}$. For both rows and columns, the procedure is repeated for a prefixed number of iterations, and the lowest-temperature candidate solution $(r^*,c^*)$ is then chosen as the final solution.
In our study, we run the BINMATNEST algorithm through the \url{nestedrank} function of the \url{bipartite} 
R package~\footnote{\url{https://www.rdocumentation.org/packages/bipartite/versions/}}.

\subsection{Fitness-complexity}

The fitness-complexity algorithm has been introduced to simultanoeusly measure the economic competitivenss of countries ($f_i\in[0,\infty)$) and the sophistication of products ($q_\alpha\in[0,\infty$) from the bipartite network connecting the countries with the products they export in world trade~\cite{tacchella2012new}.
The original fitness-complexity equations read~\cite{tacchella2012new}
\begin{equation}
\begin{split}
f_i^{-1}&=x_i=\frac{1}{\sum_a A_{ia}\,q_a} \\
q_a&=y_a=\frac{1}{\sum_i A_{ia}\,x_i},
\end{split}
\label{eq:fc}
\end{equation}
which implies that high-fitness countries export many products -- both high- and low-complexity ones -- and high-complexity products are rarely exported by low-fitness countries.
We observe that the fitness-complexity equations are formally 
equivalent to the Sinkhorn-Knopp equations used in optimal transport~\cite{sinkhorn1967concerning,mazzilli2022fitness}.
As such, they can be derived by solving a quadratic optimization problem with logarithmic barriers, defined by the energy function~\cite{marshall1968scaling}
\begin{equation}
E=\sum_{i,a}A_{ia}\,x_i\,y_a-\sum_i \log x_i-\sum_\alpha \log y_a.
\label{eq:logbarrier}
\end{equation}
By taking the partial derivatives of $E(\mathbf{x},\mathbf{y})$ with respect to $x_i$ and $y_\alpha$, respectively, we obtain indeed the fitness-complexity equations in Eq.~\eqref{eq:fc}.
This remark provides an optimization-based interpretation of the fitness-complexity equations, while it does not provide a principled interpretation for the logarithmic barriers and the relation between the fitness-complexity scores and the degree of nestedness of a network.
The algorithm has been shown to effectively pack bipartite adjacency matrices into nested configurations through both qualitative and quantitative arguments~\cite{tacchella2012new,lin2018nestedness}, which motivates its inclusion in our paper.


\subsection{Minimal extremal metric}

The minimal extremal metric (MEM) is a variant of the fitness-complexity algorithm that penalizes more heavily products exported by low-fitness countries.
The MEM equations read~\cite{wu2016mathematics}
\begin{equation}
\begin{split}
f_i^{-1}&=x_i=\frac{1}{\sum_a A_{ia}\,q_a} \\
q_a&=y_a=\min_{i:A_{ia}=1}\{F_i\},
\end{split}
\label{eq:mem}
\end{equation}
which implies high-complexity products are never exported by low-fitness countries.
The metric has been shown to visually pack bipartite adjacency matrices better than the original FC algorithm~\cite{wu2016mathematics}, which motivates its inclusion in our paper.

\section{Derivation of the saddle point equations}
\label{sec:SI_theory}
In this section we discuss in detail how to derive 
the saddle point Eqs.~\eqref{eq:sp1} given in the 
main text. We consider the minimization problem 
defined by
\begin{equation}
(r^*,s^*) = 
\argmin_{r\in\mathcal{R}_{N}, s\in \mathcal{R}_{M})}{E(r,s)}\ ,
\label{eq:si_problem}
\end{equation}
where the cost (energy) function is given by
\begin{equation}
E=\sum_{i=1}^N\sum_{a=1}^MA_{ia}\,r_i\,s_a\ ,
\label{eq:si_cost}
\end{equation}
and $\mathcal{R}_{N}$ and $\mathcal{R}_{M}$ 
are the sets of all vectors $r$ and $s$ obtained 
by permuting the entries of the representative vectors 
$r^0$ and $s^0$ defined as
\begin{equation}
\begin{aligned}
r^0 &\equiv (1,2,3,...,N)\ ,\\
s^0 &\equiv (1,2,3,...,M)\ .
\end{aligned}
\label{eq:si_representative}
\end{equation}
Therefore, we can write any two vectors $r$ and $s$ as
\begin{equation}
\begin{aligned}
r_i &= \sum_{j=1}^N P_{ij}r^0_j ,\\
s_a &= \sum_{a=1}^M Q_{ab}s^0_b\ ,
\end{aligned}
\label{eq:si_vectors}
\end{equation}
where $P$ and $Q$ are arbitrary permutation 
matrices of size $N\times N$ and $M\times M$, 
respectively.
Furthermore, we introduce the $N\times M$ matrix 
$B$ defined as the tensor product of $r^0$ and $s^0$, 
whose components are explicitly 
given by 
\begin{equation}
B_{ia} = (r^0\otimes s^0)_{ia} = ia\ .
\label{eq:si_B}
\end{equation}
With these definitions we can rewrite the energy 
function as the trace of a product of matrices 
in the following way:
\begin{equation}
E\equiv E(P,Q) = {\rm Tr}(P^tAQB^t)\ .
\label{eq:si_trace}
\end{equation}
The minimization problem in Eq.~\eqref{eq:si_problem} 
can be reformulated as a minimization problem in the 
space of permutation matrices as follows 
\begin{equation}
(P^*,Q^*)=\argmin_{(P\in \mathcal{S}_{N},\ Q\in \mathcal{S}_{M}) }{E(P,Q)}\ ,
\label{eq:si_problem2}
\end{equation}
where $\mathcal{S}_{N}$ and $\mathcal{S}_{M}$ denote the 
symmetric groups on $N$ and $M$ elements, respectively. 

Next we discuss a relaxation of the problem in 
Eq.~\eqref{eq:si_problem2} that amounts to extend 
the spaces $\mathcal{S}_{N}$ and $\mathcal{S}_{M}$ 
of permutation matrices onto the spaces of 
doubly-stochastic (DS) matrices $\mathcal{D}_N$ and 
$\mathcal{D}_M$. The space $\mathcal{D}_N$ 
($\mathcal{D}_M$) is a superset of the original space 
$\mathcal{S}_{N}$ ($\mathcal{S}_{M}$). Solving the 
problem on the $\mathcal{D}$-space means to find 
two doubly-stochastic matrices $X^*$ and $Y^*$ that 
minimize an `effective' cost function $F$, i.e. 
\begin{equation}
F(X^*,Y^*) = \min_{(X\in \mathcal{D}_{N},\ Y\in \mathcal{D}_{M}) }{F(X,Y)}\ ,
\label{eq:si_relaxed}
\end{equation}
and are only `slightly different' from the permutation 
matrices $P^*$ and $Q^*$ (we will specify later what
`slightly different' means in mathematical terms 
and what $F$ actually is).
The quantity which plays the fundamental role  
in the relaxation procedure of the original problem 
is the partition function, $Z(\beta)$, defined by
\begin{equation}
Z(\beta) = \sum_{P\in\mathcal{S}_{N}}
\sum_{Q\in\mathcal{S}_{M}}e^{-\beta E(P,Q)}\ .
\label{eq:si_partitionfunction}
\end{equation}
The connection between $Z(\beta)$ and the original 
problem in Eq.~\eqref{eq:si_problem2} is established 
by the following limit:
\begin{equation}
\lim_{\beta\to\infty}-\frac{1}{\beta}\log Z(\beta) = 
\min_{(P\in \mathcal{S}_{N},\ Q\in \mathcal{S}_{M}) }{E(P,Q)}\ .
\label{eq:si_limit}
\end{equation}
The optimization problem in Eq.~\eqref{eq:si_problem2} 
is thus equivalent to the problem of calculating the 
partition function in Eq.~\eqref{eq:si_partitionfunction}. 
Ideally, we would like to compute exactly $Z(\beta)$ 
for arbitrary $\beta$ and then take the limit 
$\beta\to\infty$. Although an exact calculation of 
the partition function is, in general, out of reach, 
in practice we may well expect that the better we
estimate $Z(\beta)$, the closer the limit in 
Eq.~\eqref{eq:si_limit} will be to the true optimal 
solution. 
In fact, the procedure of relaxation is basically 
a procedure to assess the partition function for 
large but finite $\beta$. Mathematically, this 
procedure is called method of steepest descent~\cite{debye}. 
By estimating the partition function via the 
steepest descent method we will obtain a system 
of non-linear equations, called saddle-point equations, 
whose solution is a pair of doubly-stochastic 
matrices $X^*,Y^*$ that solve the relaxed problem 
given by Eq.~\eqref{eq:si_relaxed}. 
Eventually, the solution to the original problem in 
Eq.~\eqref{eq:si_problem2} can be obtained formally by 
projecting $X^*,Y^*$ onto the subspaces 
$\mathcal{S}_{N},\mathcal{S}_{M}\subset 
\mathcal{D}_{N},\mathcal{D}_{M}$ via the limit
\begin{equation}
\begin{aligned}
\lim_{\beta\to\infty} X^*(\beta) &= P^*\ ,\\
\lim_{\beta\to\infty} Y^*(\beta) &= Q^*\ .
\end{aligned}
\label{eq:si_projection}
\end{equation}
Having explained the rationale for the introduction 
of the partition function, we move next to discuss 
the details of the calculation leading to the saddle 
point equations. 


In order to cast the partition function 
in a form suitable for the steepest-descent 
evaluation, we need the following preliminary 
result. 

{\bf Definition: Semi-permutation matrix:} 
a $N\times N$ square matrix $\slashed{P}$ is called a 
semi-permutation matrix if $\slashed{P}_{ij}\in\{0,1\}$ 
and each row sums to one, i.e. 
$\sum_{j=1}^N\slashed{P}_{ij}=1$ for $i=1,...,N$, 
but no further constraint on the column sums is 
imposed. 

\bigskip

We denote $\mathcal{\slashed{S}}_N$ the space of 
semi-permutation matrices:
\begin{equation}
\mathcal{\slashed{S}}_N = \Bigg\{
\slashed{P}\ |\ \slashed{P}_{ij}\in\{0,1\}\ {\rm AND\ }
\sum_{j=1}^N\slashed{P}_{ij}=1\  \forall i\Bigg\}
\end{equation}


{\bf Lemma}

Consider an arbitary $N\times N$ square matrix $G$ 
and the function $W(G)$ defined by
\begin{equation}
e^{W(G)} = \sum_{\slashed{P}\in \mathcal{\slashed{S}}_N}
e^{{\rm Tr}(\slashed{P}G^t) }\ .
\label{eq:si_expW}
\end{equation}
Then, $W(G)$ is explicitly given by the following formula
\begin{equation}
\boxed{\ 
W(G) = \sum_{i=1}^N\log\sum_{j=1}^N e^{G_{ij}}\ }\ .
\label{eq:si_W}
\end{equation}


{\bf Proof}

Let us write the right hand side of Eq.~\eqref{eq:si_expW} 
as
\begin{equation}
\sum_{\slashed{P}\in \mathcal{\slashed{S}}_N}
e^{\sum_{ij}\slashed{P}_{ij}G_{ij}} = 
\sum_{\slashed{P}_1}e^{\sum_{j}(\slashed{P}_1)_jG_{1j}} 
\sum_{\slashed{P}_2}e^{\sum_{j}(\slashed{P}_2)_jG_{2j}}
\ \ldots\ ,
\label{eq:si_proof1}
\end{equation}
where $\slashed{P}_i$ is the ${\rm i}^{\rm th}$ 
row of $\slashed{P}$ (and thus is a vector) having 
one component equal to $1$ and the remaining $N-1$ 
components equal to $0$. The sum $\sum_{\slashed{P}_i}$ 
denotes a summation over all possible choices of 
the vector $\slashed{P}_i$: there are $N$ possible 
such choices, namely $\slashed{P}_i=(1,0,...,0)$, 
$\slashed{P}_i=(0,1,...,0), ..., \slashed{P}_i=(0,0,...,1)$. 
Hence, each sum in the right hand side of 
Eq.~\eqref{eq:si_proof1} evaluates
\begin{equation}
\sum_{\slashed{P}_i}e^{\sum_{j}(\slashed{P}_i)_jG_{ij}} = 
e^{G_{i1}} + e^{G_{i2}} +... = 
\sum_{j=1}^Ne^{G_{ij}}\ . 
\end{equation}
Thus, the left hand side of Eq.~\eqref{eq:si_proof1} 
is equal to 
\begin{equation}
\sum_{\slashed{P}\in \mathcal{\slashed{S}}_N}
e^{\sum_{ij}\slashed{P}_{ij}G_{ij}} = 
\prod_{i=1}^N \sum_{j=1}^Ne^{G_{ij}}\ .
\label{eq:si_proof2}
\end{equation}
Eventually, by taking the logarithm of both sides 
of Eq.~\eqref{eq:si_proof2}, we prove Eq.~\eqref{eq:si_W}.

With these tools at hand we move to derive the integral 
representation of $Z(\beta)$.

\subsection*{Integral representation of $Z(\beta)$}
We use the definition of the Dirac $\delta$-function to 
write the partition function in Eq.~\eqref{eq:si_partitionfunction} 
as follows
\begin{equation}
Z(\beta) = \sum_{P\in\mathcal{S}_{N}}
\sum_{Q\in\mathcal{S}_{M}}\int DX\int DY 
e^{-\beta E(X,Y)}
\prod_{i,j=1}^N\delta(X_{ij}-P_{ij})
\prod_{a,b=1}^N\delta(Y_{ab}-Q_{ab})\ ,
\label{eq:si_partitionfunction2}
\end{equation}
where the integration measures are defined by 
$DX\equiv\prod_{i,j}dX_{ij}$ and 
$DY\equiv\prod_{a,b}dY_{ab}$. 
The next step is to transform the sum over 
permutation matrices $P,Q$ into a sum over 
semi-permutations matrices $\slashed{P}, \slashed{Q}$ 
and then performing explicitly this sum using 
the Lemma in Eq.~\eqref{eq:si_W}. 
In order to achieve this goal, we insert into 
Eq.~\eqref{eq:si_partitionfunction2} $N$ delta 
functions $\prod_{j=1}^N \delta\Big(\sum_iX_{ij}-1\Big)$ 
and $M$ delta functions 
$\prod_{b=1}^M \delta\Big(\sum_aY_{ab}-1\Big)$ 
to enforce the conditions that the columns of 
$X$ and $Y$ do sum up to one. By inserting these
delta functions, we can then replace the sum 
over $P,Q$ by a sum over $\slashed{P}, \slashed{Q}$, 
thus obtaining 
\begin{equation}
Z(\beta) = \sum_{\slashed{P}}
\sum_{\slashed{Q}}
\int DXDY 
e^{-\beta E(X,Y)}
\prod_{i,j=1}^N\delta(X_{ij}-\slashed{P}_{ij})
\prod_{a,b=1}^N\delta(Y_{ab}-\slashed{Q}_{ab})
\prod_{j=1}^N \delta\Big(\sum_iX_{ij}-1\Big)
\prod_{b=1}^M \delta\Big(\sum_aY_{ab}-1\Big)\ .
\label{eq:si_partitionfunction3}
\end{equation}
To proceed further in the calculation, we use 
the following integral representations of the 
delta-functions:
\begin{equation}
\begin{aligned}
\delta(X_{ij} - \slashed{P}_{ij}) &= 
\frac{1}{2\pi i}\int_{-i\infty}^{i\infty} 
d\hat{X}_{ij}\ e^{-\hat{X}_{ij}(X_{ij}-\slashed{P}_{ij})}\ ,\\
\delta(Y_{ab} - \slashed{Q}_{ab}) &= 
\frac{1}{2\pi i}\int_{-i\infty}^{i\infty} 
d\hat{Y}_{ab}\ e^{-\hat{Y}_{ab}(Y_{ab}-\slashed{Q}_{ab})}\ ,\\
\delta\Big(\sum_iX_{ij}-1\Big) &= 
\frac{1}{2\pi i}\int_{-i\infty}^{i\infty} 
dz_j\ e^{-z_j\big(\sum_iX_{ij}-1\big)}\ ,\\
\delta\Big(\sum_aY_{ab}-1\Big) &= 
\frac{1}{2\pi i}\int_{-i\infty}^{i\infty} 
dw_b\ e^{-w_b\big(\sum_aY_{ab}-1\big)}\ ,
\end{aligned}
\end{equation}
into Eq.~\eqref{eq:si_partitionfunction3} and we get 
\begin{equation}
\begin{aligned}
Z(\beta) = \sum_{\slashed{P}}
\sum_{\slashed{Q}}\int DXDYD\hat{X}D\hat{Y}DzDw\  
&e^{-\beta E(X,Y)}
e^{-{\rm Tr}(\hat{X}X^t) + {\rm Tr}(\hat{X}\slashed{P}^t) 
-{\rm Tr}(\hat{Y}Y^t) + {\rm Tr}(\hat{Y}\slashed{Q}^t)}
\times \\
&\times e^{-\sum_jz_j\big(\sum_iX_{ij}-1\big)}
e^{-\sum_bw_b\big(\sum_aY_{ab}-1\big)}\ ,
\end{aligned}
\label{eq:si_partitionfunction4}
\end{equation}
where we defined the integration measures 
$D\hat{X}\equiv\prod_{i,j}d\hat{X}_{ij}/2\pi i$, 
$D\hat{Y}\equiv\prod_{a,b}d\hat{Y}_{ab}/2\pi i$,  
$Dz\equiv\prod_{j}dz_{j}/2\pi i$, and 
$Dw\equiv\prod_{b}dw_{b}/2\pi i$. 
Performing the sums over $\slashed{P}$ and $\slashed{Q}$ 
using Eq.~\eqref{eq:si_W} we obtain
\begin{equation}
\begin{aligned}
Z(\beta) = \int DXDYD\hat{X}D\hat{Y}DzDw\  
&e^{-\beta E(X,Y)}
e^{-{\rm Tr}(\hat{X}X^t) + W(\hat{X}) 
-{\rm Tr}(\hat{Y}Y^t) + W(\hat{Y})}
\times \\
&\times e^{-\sum_jz_j\big(\sum_iX_{ij}-1\big)}
e^{-\sum_bw_b\big(\sum_aY_{ab}-1\big)}\ .
\end{aligned}
\label{eq:si_partitionfunction5}
\end{equation}
Next we introduce the {\bf effective cost function} 
$F(X,\hat{X},Y,\hat{Y},z,w)$ defined as 
\begin{equation}
\begin{aligned}
F(X,\hat{X},Y,\hat{Y},z,w) &= E(X,Y) +
\frac{1}{\beta}{\rm Tr}(\hat{X}X^t) +
\frac{1}{\beta}{\rm Tr}(\hat{Y}Y^t) -
\frac{1}{\beta}W(\hat{X}) -
\frac{1}{\beta}W(\hat{Y}) +\\
&+  
\frac{1}{\beta}\sum_jz_j\big(\sum_iX_{ij}-1\big) +
\frac{1}{\beta}\sum_bw_b\big(\sum_aY_{ab}-1\big) \equiv\\
&\equiv E(X,Y) -\frac{1}{\beta}S(X,\hat{X},Y,\hat{Y},z,w)
\end{aligned}
\label{eq:si_fenergy}
\end{equation}
whereby we can write the partition function as 
\begin{equation}
Z(\beta) = \int DXDYD\hat{X}D\hat{Y}DzDw\ 
e^{-\beta F(X,\hat{X},Y,\hat{Y},z,w)}\ ,
\label{eq:si_partitionfunction6}
\end{equation}
which can be evaluated by the steepest descent 
method when $\beta\to\infty$, as we explain next.

\subsection*{Steepest descent evaluation of the partition function}
In the limit of large $\beta$ the integral in 
Eq.~\eqref{eq:si_partitionfunction6} is dominated 
by the saddle point where $E(X,Y)$ is minimized 
and $S(X,\hat{X},Y,\hat{Y},z,w)$ is stationary (in 
order for the oscillating contributions to not cancel 
out). In order to find the saddle point, we have to 
set the derivatives of $F(X,\hat{X},Y,\hat{Y},z,w)$ 
to zero, thus obtaining the following {\bf saddle 
point equations}
\begin{equation}
\begin{aligned}
\frac{\partial F}{\partial X_{ij}} &= 
\frac{\partial E}{\partial X_{ij}} +
\frac{1}{\beta}\big(\hat{X}_{ij}+z_j\big) = 0\ , \\
\frac{\partial F}{\partial \hat{X}_{ij}} &= 
\frac{1}{\beta}X_{ij}
-\frac{1}{\beta}\frac{\partial W}{\partial \hat{X}_{ij}}\ ,\\
\frac{\partial F}{\partial z_j} &= \sum_i X_{ij}-1 = 0\ ,
\end{aligned}
\label{eq:si_saddlepoint}
\end{equation}
and similar equations for the triplet $(Y,\hat{Y},w)$. 
The derivative of $E$ with respect to $X_{ij}$ gives 
\begin{equation}
\frac{\partial E}{\partial X_{ij}} = (AYB^t)_{ij}\ , 
\end{equation}
and the derivative of $W$ with respect to $\hat{X}_{ij}$ 
gives
\begin{equation}
\frac{\partial W}{\partial \hat{X}_{ij}} = 
\frac{ e^{\hat{X}_{ij}} }{\sum_k e^{\hat{X}_{ik}}}\ .
\end{equation}
Solving Eq.~\eqref{eq:si_saddlepoint} with respect 
to $X_{ij}$ we get
\begin{equation}
X_{ij} = \frac{e^{-\beta(AYB^t)_{ij} - z_j }}
{\sum_k e^{-\beta(AYB^t)_{ik} - z_k }}\ .
\label{eq:si_solutionX}
\end{equation}
Analogously, solving with respect to $Y_{ab}$ 
we get
\begin{equation}
Y_{ab} = \frac{e^{-\beta(A^tXB)_{ab} - w_b }}
{\sum_c e^{-\beta(A^tXB)_{ac} - w_c }}\ .
\label{eq:si_solutionY}
\end{equation}
It is worth noticing that Eqs.~\eqref{eq:si_solutionX} and~\eqref{eq:si_solutionY} are invariant under the 
tranformations
\begin{equation}
\begin{aligned}
z_j\ &\to\ z_j + \zeta\ ,\\
w_b\ &\to\ w_b + \xi\ ,
\end{aligned}
\label{eq:si_symmetry}
\end{equation}
for arbitrary values of $\zeta$ and $\xi$. 
This translational symmetry is due to the 
fact that the $2N$ constraints on the row and 
column sums of $P$ are not linearly independent, 
since the sum of all entries of $P$ must be equal 
to $N$, i.e. $\sum_{ij}P_{ij}=N$. 
The same reasoning applies to the $2M$ constraints 
on the row and column sums of $Q$, of which only 
$2M-1$ are linearly independent, since $\sum_{ab}Q_{ab}=M$.
Furthermore, we notice that the solutions matrices 
$X$ and $Y$ in Eqs.~\eqref{eq:si_solutionX},~\eqref{eq:si_solutionY} 
automatically satisfy the condition of having row 
sums equal to one.
Next, we derive the equations to determine the 
Lagrange multipliers $z_j$ and $w_b$. 
To this end we first introduce the vectors $v$ 
and $\nu$ with components
\begin{equation}
\begin{aligned}
v_j &= e^{-z_j}\ ,\\
\nu_b &= e^{-w_b}\ .
\label{eq:si_rightvec}
\end{aligned}
\end{equation}
Then, we define the vectors $u$ and $\mu$ as 
\begin{equation}
\begin{aligned}
u_i   &= \Big(\sum_k e^{-\beta(AYB^t)_{ik}}\ v_k\Big)^{-1}\ ,\\
\mu_a &= \Big(\sum_c e^{-\beta(A^tXB)_{ac}}\ \nu_c\Big)^{-1}\ ,
\label{eq:si_leftvec}
\end{aligned}
\end{equation}
so that we can write the solutions matrices $X$ and $Y$ 
in Eqs.~\eqref{eq:si_solutionX},~\eqref{eq:si_solutionY} 
as
\begin{equation}
\begin{aligned}
X_{ij} &= u_i\  e^{-\beta(AYB^t)_{ij}}\ v_j,\\
Y_{ab} &= \mu_a\ e^{-\beta(A^tXB)_{ab}}\ \nu_b\ .
\end{aligned}
\label{eq:si_solutionXY}
\end{equation}
Finally, imposing the conditions on $X$ and $Y$ 
to have column sums equal to one, we find the 
equations to be satisfied by $v$ and $\nu$
\begin{equation}
\begin{aligned}
v_j   &= \Big(\sum_i u_i\  e^{-\beta(AYB^t)_{ij}}\Big)^{-1}\ ,\\
\nu_b &= \Big(\sum_a \mu_a e^{-\beta(A^tXB)_{ab}}\Big)^{-1}\ ,
\label{eq:si_rightvec2}
\end{aligned}
\end{equation}
Equations~\eqref{eq:si_leftvec},~\eqref{eq:si_solutionXY}, 
and~\eqref{eq:si_rightvec2} are the constitutive 
equations for the relaxed nestedness-maximization 
problem corresponding to Eqs.~\eqref{eq:sp1} given 
in the main text. 

We conclude this section by deriving the self-consistent 
equations for the ``stochastic rankings'' corresponding 
to Eqs.~\eqref{eq:rankPropagation1} and~\eqref{eq:rankPropagation2} 
given in the main text. 
We define the stochastic rankings as the two vectors 
\begin{equation}
\begin{aligned}
\rho_i   &= \sum_{k=1}^NX_{ik}\ k\ ,\\
\sigma_a &= \sum_{a=1}^M Y_{ab}\ b\ ,
\end{aligned}
\end{equation}
where the term ``stochastic'' emphasizes their 
implied dependence on the doubly stochastic matrices 
$X$ and $Y$. Clearly we have
\begin{equation}
\begin{aligned}
\lim_{\beta\to\infty}\rho_i &= r_i\ ,\\
\lim_{\beta\to\infty}\sigma_a &= s_a\ .
\end{aligned}    
\end{equation}
Next, let's consider the argument of the exponentials 
in Eq.~\eqref{eq:si_solutionXY}, that we can rewrite as 
\begin{equation}
\begin{aligned}
(AYB^t)_{ij} &= \sum_a A_{ia}\Big( \sum_b Y_{ab}\ b\Big) j = 
j\sum_a A_{ia}\sigma_a,\\
(A^tXB)_{ab} &= \sum_i A_{ia}\Big(\sum_j X_{ij}\ j\Big) b = 
b\sum_i A_{ia} \rho_i\ .
\end{aligned}
\end{equation}
At this point is sufficient to multiply both sides 
of Eq.~\eqref{eq:si_solutionXY} by $j$ and 
$b$, and sum over $j$ and $b$, respectively, 
to obtain 
\begin{equation}
\begin{aligned}
\sum_jX_{ij}\ j &= \rho_i = u_i\sum_je^{-\beta(AYB^t)_{ij}}\ v_j\ j=
u_i\sum_je^{-\beta j\sum_a A_{ia}\sigma_a}\ v_j\ j\ ,\\
\sum_b Y_{ab}\ b &= \sigma_a = 
\mu_a\sum_be^{-\beta(A^tXB)_{ab}}\ \nu_b\ b = 
\mu_a\sum_be^{-\beta b\sum_i A_{ia} \rho_i}\ \nu_b\ b\ .
\end{aligned}
\label{eq:si_solution_rs}
\end{equation}
Using the definition of $u_i$ and $\mu_a$ in 
Eqs.~\eqref{eq:si_leftvec} we obtain
\begin{equation}
\begin{aligned}
\rho_i &= \frac{\sum_je^{-\beta j\sum_a A_{ia}\sigma_a}\ v_j\ j}{\sum_je^{-\beta j\sum_a A_{ia}\sigma_a}\ v_j}\ ,\\
\sigma_a &= 
\frac{\sum_be^{-\beta b\sum_i A_{ia} \rho_i}\ \nu_b\ b}{\sum_be^{-\beta b\sum_i A_{ia} \rho_i}\ \nu_b}\ ,
\end{aligned}
\label{eq:si_self_rs}
\end{equation}
which are the self-consistent 
Eqs.~\eqref{eq:rankPropagation1} for $\rho$ and $\sigma$ 
given in the main text. 
There are still two unknown vectors in the previous 
equations: vectors $v$ and $\nu$. In order 
to determine them we consider Eqs.~\eqref{eq:si_rightvec2} 
and eliminate $u_i$ and $\mu_a$ using Eqs.~\eqref{eq:si_leftvec}, 
thus obtaining 
\begin{equation}
\begin{aligned}
v_j = \Bigg(
\sum_i\Big[\sum_k\ v_k\   e^{-\beta (k-j)\sum_a A_{ia}\sigma_a}\Big]^{-1}\Bigg)^{-1} ,\\ 
\nu_b = \Bigg(
\sum_a\Big[\sum_c\ \nu_c\ e^{-\beta (c-b)\sum_i A_{ia}\rho_i}\Big]^{-1}
\Bigg)^{-1}\ , 
\end{aligned}
\label{eq:si_self_vnu}
\end{equation}
which are the self-consistent Eqs.~\eqref{eq:rankPropagation2} 
for $v$ and $\nu$ given in the main text. 

In the next section we describe a simple iterative 
algorithm to solve Eqs.~\eqref{eq:si_self_rs} 
and~\eqref{eq:si_self_vnu}.

\section{Algorithm}
\label{sec:algorithm}
The algorithm to solve Eqs.~\eqref{eq:si_self_rs} 
and~\eqref{eq:si_self_vnu} consists of 4 basic steps, 
explained below.
\begin{enumerate}
\item Initialize $\rho_i$ uniformly at random in $[1,N]$; 
similarly, initialize $\sigma_a$ uniformly at random in $[1,M]$.
Also, initialize $v_j$ and $\nu_b$ uniformly at random in $(0,1]$.  
\item Choose an initial value for $\beta$. To start, initialize 
$\beta$ using the following formula:
\begin{equation} 
\beta = \beta_{\rm init} = \frac{1}{\max\Big[N \max_i\{k_i\}, M \max_a\{k_a\}\Big]}\ ,
\end{equation}
where $k_i = \sum_a A_{ia}$, and $k_a = \sum_i A_{ia}$. 
\item Set $\tau=1$, and a tolerance ${\rm TOL} = 10^{-3}$. 
Then run the following subroutine. 
\begin{enumerate}
\item Iterate Eqs.~\eqref{eq:si_self_vnu} according to the following 
updating rules
\begin{equation}
\begin{aligned}
v_j(t+1) = \Bigg(
\sum_i\Big[\sum_k\ v_k(t)\   e^{-\beta (k-j)\sum_a A_{ia}\sigma_a}\Big]^{-1}\Bigg)^{-1} ,\\ 
\nu_b(t+1) = \Bigg(
\sum_a\Big[\sum_c\ \nu_c(t)\ e^{-\beta (c-b)\sum_i A_{ia}\rho_i}\Big]^{-1}
\Bigg)^{-1}\ , 
\end{aligned}
\label{eq:si_self_vnu_iterative}
\end{equation}
until convergence. 
\item Iterate Eqs.~\eqref{eq:si_self_rs} according to the following 
updating rules
\begin{equation}
\begin{aligned}
\rho_i(t+1) &= \frac{\sum_je^{-\beta j\sum_a A_{ia}\sigma_a(t)}\ v_j\ j}{\sum_je^{-\beta j\sum_a A_{ia}\sigma_a(t)}\ v_j}\ ,\\
\sigma_a(t+1) &= 
\frac{\sum_be^{-\beta b\sum_i A_{ia} \rho_i(t)}\ \nu_b\ b}{\sum_be^{-\beta b\sum_i A_{ia} \rho_i(t)}\ \nu_b}\ ,
\end{aligned}
\label{eq:si_self_rs_iterative}
\end{equation}
until convergence. Call $\rho_i^{(\tau)}$ and $\sigma_a^{(\tau)}$ 
the converged vectors and compute 
\begin{equation}
{\rm MAXDIFF}\equiv\max\Bigg\{\max_i\Big[\rho_i^{(\tau)} - \rho_i^{(\tau-1)}\Big], 
    \max_a\Big[\sigma_a^{(\tau)} - \sigma_a^{(\tau-1)}\Big]\Bigg\} .
\end{equation}
\item If ${\rm MAXDIFF} < {\rm TOL}$, then RETURN $\rho_i^{(\tau)}$ 
and $\sigma_a^{(\tau)}$; otherwise increase $\tau$ by 1 and repeat 
from (a).
\end{enumerate}
\item Increase $\beta\to \beta + d\beta$ and repeat from (3) or 
terminate if the returned vectors did not change from the 
iteration at $\beta - d\beta$.
\end{enumerate}

Having found the solution vectors $\rho$ and $\sigma$, we convert 
them into integer rankings as follows. The smallest value of $\rho_i$ 
is assigned rank 1. The second smallest is assigned rank 2, and so 
on and so forth. This procedure generates a mapping from $1,2,...,N$ 
to $i_1,i_2,...,i_N$ that can be represented by a $N\times N$ 
permutation matrix $P_{ij}$. The same procedure, applied to $\sigma_a$, 
generates a $M\times M$ permutation matrix $Q_{ij}$. Matrices $P$ 
and $Q$ represent the optimal permutations that solve the nestedness 
maximization problem. 
Eventually, application of the similarity transformation 
\begin{equation}
A\to P^tAQ\ ,
\end{equation}
brings the adjacency matrix into its maximally nested form 
having all nonzero entries clustered in the upper left 
corner, as seen in Fig.~\ref{fig:fig2}c.

%


\end{document}